\documentclass[aps,pra,amsmath,amssymb,floatfix,twocolumn,superscriptaddress]{revtex4}
\usepackage{graphicx}

\newcommand{\exv}[3]{\left\langle{#1}\right\vert{#2}\left\vert{#3}\right\rangle}
\newcommand{\Ra}{{\bf R}_A}

\newcommand{\dr}{\,d{\bf r}} 
\newcommand{\abs}[1]{\left|#1\right|}
\newcommand{\hal}{\frac{1}{2}} 

\newcommand{\pa}{\partial} 
\newcommand{\ro}{{\bf r}} 
\newcommand{\rb}{({\bf r})} 
\newcommand{\rp}{{\bf r'}} 
\newcommand{\rpb}{({\bf r'})}
\newcommand{\rtb}{({\bf r}\,t)} 
\newcommand{\rptb}{({\bf r'} t)}
\newcommand{\drp}{\,d{\bf r'}} 
\newcommand{\dintp}{\int\!\!\int'}
\begin{document}
\title{ Importance of electronic self-consistency in the TDDFT based treatment
of nonadiabatic molecular dynamics } 
\author{T.A. Niehaus} 
\author{D. Heringer} 
\affiliation{ Dept. of Theoretical Physics, University of Paderborn, %
 D - 33098 Paderborn, Germany } 
\author{B. Torralva}
\affiliation{Chemistry and Materials Science, Lawrence Livermore National %
 Laboratory, Livermore, CA 94550, USA}
\author{Th. Frauenheim } 
\affiliation{ Dept. of Theoretical Physics, University of Paderborn, %
 D - 33098 Paderborn, Germany } 
\date{\today}
\begin{abstract}
A mixed quantum-classical approach to simulate the coupled dynamics of
electrons and nuclei in nanoscale molecular systems is presented. The method
relies on a second order expansion of the Lagrangian in time-dependent density
functional theory (TDDFT) around a suitable reference density. We show that
the inclusion of the second order term renders the method a self-consistent
scheme and improves the calculated optical spectra of molecules by a proper
treatment of the coupled response. In the application to ion-fullerene
collisions, the inclusion of self-consistency is found to be crucial for a
correct description of the charge transfer between projectile and target.  For
a model of the photoreceptor in retinal proteins, nonadiabatic molecular
dynamics simulations are performed and reveal problems of TDDFT in the
prediction of intra-molecular charge transfer excitations.
\end{abstract}

\maketitle

\section{Introduction}
Beginning with the work of Zangwill and Soven \cite{Zan80}, the generalization
of density functional theory to time dependent phenomena (TDDFT) has become an
important tool in the description of laser-matter interaction.  The possible
applications are diverse and range from the calculation of spectra (linear
optical \cite{Bau96,Vas99,Deb01}, circular dichroism \cite{Fur00,Yab99},
resonant Raman \cite{Her04}) to the evaluation of properties (polarization
\cite{Van961,Van962}, hyperpolarization \cite{Van97}) up to studies of high
harmonic generation \cite{Ull96,Chu01,Cas04} and photochemical reactions
\cite{Dou04}. The formal justification of TDDFT was laid by Runge and Gross
\cite{Run84}, who showed that the exact many body electron density can be
obtained from single-particle mean field equations. The solution of these time
dependent Kohn-Sham (TDKS) equations can be obtained either perturbatively in
the small amplitude limit \cite{Cas95} or by direct numerical integration in
the time domain \cite{Yab96}. Both approaches have their inherent merits and
disadvantages.

In the linear response regime, for example, the problem can be recast in an
eigenvalue equation in the particle-hole representation. This allows for an
interpretation of optical spectra in terms of contributing single-particle
transitions and also for a symmetry assignment of the states. Moreover,
transitions with vanishing oscillator strength can be located, like dark
singlet or generally triplet states \cite{Com3}. One of the drawbacks of this
approach is the rather poor numerical performance with a scaling of $N^6$ ,
where N is the number of electrons. It should be noted, however, that the CPU
time as well as the memory demand can be significantly reduced when iterative
procedures like the Davidson algorithm are employed.

In terms of scaling behavior, the numerical integration of the TDKS equations
is much more favorable. Here, only the set of occupied orbitals needs to be
treated. Because the propagation involves only matrix-vector products, linear
scaling can be achieved for large systems \cite{Yam03}. Moreover, since this
approach is not restricted to small intensities, non-linear effects like
harmonic generation or multiphoton processes can be addressed. Another
advantage of working in the real time domain is the possibility to study
systematically the effect of different pulse shapes of the laser field on
observables such as ionization \cite{Cal99}. With todays femtosecond laser
sources, this is currently an active field of experimental research
\cite{Bri01}.

Nearly all first principles applications of the real time approach have been
limited to systems with fixed nuclei. Clearly, it would be highly desirable to
study the motion of the coupled system of electrons and nuclei, which would
allow one to address problems like laser induced vibrational excitation or
photochemical reactions. Since the time step of such molecular dynamics
simulations is set to attoseconds by the ultrafast electronic motion, only
small systems with a few degrees of freedom can be treated in an ab initio
frame work \cite{Cas04,Cal99}. Consequently, approximate TDDFT methods are
quite successful in this domain of application.

 Different groups contributed to this field and used their developments in a
variety of different
studies \cite{Dou04,All94,Tor01,Tor02,Dou031,Dou032,Dou033,Dou034,Saa96,Saa98,%
Sch97,Kno99,Kno00,Kun01,Kun03,Tod01,mydiss,Fra02}.  In all these approximate
schemes only the valence electrons are treated explicitely and the TDKS
orbitals are expanded in a limited (usually minimal) basis of atomic
orbitals. The Lagrangian, which is a functional of the time dependent density,
is then expanded around a static reference density up to a certain order. In
zeroth order the Hamiltonian depends only on the reference density, which
permits the calculation of the necessary matrix elements once and for all. In
this respect, the methods are similar to tight-binding approaches, although no
fitting to experimental data is performed.

The purpose of this work is to analyze the implications of extending the
mentioned expansion, since all studies so far were restricted to zeroth order.
After a more detailed description of the problem in Sec.~\ref{meth}, we test
the extension in the determination of optical spectra in Sec.~\ref{butasec} as
well as for nonadiabatic molecular dynamics in Sec.~\ref{coll}. Finally, we
perform an investigation of the photochemical reaction of a retinal analogue,
a chromophore which exhibits an ultrafast radiationless deactivation in
nature. These applications in quite different areas of molecular physics are
intended to investigate the transferability of the method and also to
illustrate the possibilities offered by an approximate solution of the TDDFT
equations.  

\section{Method}
\label{meth}
\enlargethispage{2cm} In order to study the dynamics of a coupled system of
electrons and nuclei, the equations of motion (EOM) need to be
determined. While the electronic EOM in the framework of DFT is given by the
well known time dependent Kohn-Sham equations, the nuclear EOM or force
equation is not a priori evident. It can be derived either by exploiting the
fact that the total energy is a conserved quantity, or by applying the
Lagrange formalism. We follow the latter approach here and define the
following Lagrangian, which depends on the TDKS states $\Psi_i\rtb$ and the
nuclear positions $\Ra$:
 \begin{eqnarray}
\label{lagra}
    {\cal L} &=& \sum_A \frac{1}{2} M_A \dot{\bf R}_A^2 \\&& -\;
   \sum_i^{\text{occ}} \exv{\Psi_i\rtb}{H[\rho]\rtb-i \frac{\pa}{\pa
   t}}{\Psi_i\rtb}- E_{\mathrm{DC}} - E_{ii},\nonumber
\end{eqnarray}     
with $\rho\rtb=\sum_i |\Psi_i\rtb|^2$.
Here the first term is the classical kinetic energy of the ions, while the
remaining terms in Eq.~(\ref{lagra}) can be obtained from the TDDFT action
functional under the assumption that the exchange-correlation (xc)
contributions are local in time \cite{Gro90}. In this widely used adiabatic
local density approximation, standard ground state functionals can also be
used in the time dependent context simply by evaluation at the time dependent
density. Thus, the Hamiltonian $H[\rho]\rtb$ in Eq.~(\ref{lagra}) takes the
common DFT form. Furthermore, $E_{\mathrm{DC}}$ represents the double counting
terms
 \begin{equation}
 E_{\mathrm{DC}}= - \hal \dintp \frac{\rho\rtb \rho\rptb}{\abs{\ro-\rp}} +
  E_{xc}[\rho] - \int v_{xc}[\rho] \rho\rtb,
\end{equation}  
and $ E_{ii}$ the ion-ion repulsion (Here and in the following $\int\!\drp$ is
abbreviated as $\int'$, and $\int\!\dr$ as $\int$).

We now proceed by applying the same kind of approximations as were used in the
derivation of the density functional theory based tight-binding (DFTB) method
\cite{Por95,Els98} from static DFT. To keep the presentation concise we refer
to some reviews \cite{Fra00,Fra02}, which provide a more detailed description
of the basic concepts, practical realization and accuracy of the ground state
DFTB approach. Here, we only report aspects, which are specific for the
generalization to the time-dependent case. In a first step, the Lagrangian is
expanded around a reference density $\rho_0(\ro)$,
$\rho\rtb=\rho_0(\ro)+\delta \rho\rtb$, which is given as a superposition of
atomic (ground state) densities. In contrast to our earlier work \cite{Fra02},
we now include terms up to second order in the density fluctuations $\delta
\rho\rtb$:
\begin{subequations}
\begin{align}
   &{\cal L}\approx \sum_A \frac{1}{2} M_A \dot{\bf R}_A^2 -
   \sum_i^{\text{occ}} \exv{\Psi_i\rtb}{H[\rho_0]\rb-i\frac{\pa}{\pa
   t}}{\Psi_i\rtb}
  \label{LAGRAl1}\\&+ \hal  \dintp
  \frac{\rho_0\rb \rho_0\rpb}{\abs{\ro-\rp}} - E_{xc}[\rho_0] + \int
  v_{xc}[\rho_0] \rho_0\rb - E_{ii} \label{LAGRAl2} \\ &- \hal \dintp \left(
  \frac{1}{\abs{\ro-\rp}} + \frac{\delta v_{xc}[\rho]\rtb}{\delta\rho\rptb }
  \right) \delta\rho\rtb \delta\rho\rptb.  \label{LAGRAl3}
\end{align}
\end{subequations}
Please note that in this expansion all contributions which are linear in $\delta
\rho$ are captured by the second term in Eq.~(\ref{LAGRAl1}) through the TDKS
states. 
The terms in Eq.~(\ref{LAGRAl2}) can now be subsumed as $E_{\text{rep}}$, a
sum of short ranged pair potentials, which depend only on the atomic species
and the interatomic distance. Since $E_{\text{rep}}$ is a functional of the
time independent reference density $\rho_0$ only, it is exactly the same as
used in the ground state DFTB scheme. The second order term of
Eq.~(\ref{LAGRAl3}), which is the focus of this work, is approximated as
follows:
\begin{eqnarray}
  \label{scc}
  E_{\text{2nd}} &=& \hal \dintp \left( \frac{1}{\abs{\ro-\rp}} + \frac{\delta
  v_{xc}[\rho]\rtb}{\delta\rho\rptb } \right) \delta\rho\rtb
  \delta\rho\rptb\nonumber\\ &\approx& \hal \sum_{AB} \Delta q_A(t)
  \gamma_{AB} \Delta q_B(t).
\end{eqnarray}
Here the $\Delta q_A(t)$ denote atomic net Mulliken charges
\begin{eqnarray}
  \label{mull}
  \Delta q_A(t) &=& q_A(t) - q_A^{\text{free atom}}\\ q_A(t) &=&
      \hal\sum_i^{\text{occ}} \sum_{\mu \in A, \nu} \left( b_{\mu i}^*(t)
      b_{\nu i}(t) S_{\mu\nu} + b_{\nu i}^*(t) b_{\mu i}(t)S_{\nu\mu}
      \right)\nonumber,
\end{eqnarray}
where the coefficients ${b_{\mu i}(t)}$ are defined by the expansion of the
TDKS states in a basis of non-orthogonal atomic orbitals $\phi_{\mu}(\ro-\Ra)$
\begin{equation}
  \label{basis}
  \Psi_{i}\rtb = \sum_{\mu} b_{\mu i}(t) \phi_{\mu}(\ro-\Ra),
\end{equation}
which build the overlap matrix $S_{\mu\nu} = \langle\phi_{\mu} |
\phi_{\nu}\rangle$.  Further, the function $\gamma_{AB}$ in Eq.~\ref{scc}
interpolates between a pure Coulomb interaction for large interatomic
distances and an element-specific constant in the atomic limit. This
numerically evaluated parameter includes the effects of exchange and
correlation and is directly related to the chemical hardness of the atomic
species \cite{Els98}. Taking the coefficients and nuclear positions as
generalized coordinates, the evaluation of the Euler-Lagrange equations leads
to the desired equations of motion. The electronic motion obeys:
\begin{subequations}
  \begin{align}
\label{SCCb}
 \dot{b}_{\nu i} &= - \sideset{}{_{\delta\mu}}{\sum\nolimits}
   S^{-1}_{\nu\delta}\left[ i H_{\delta\mu} + \sum_A \dot{\bf R}_A
   \langle\phi_\delta|\frac{\pa}{\pa \Ra} \phi_\mu\rangle\right] b_{\mu i}
   \\\intertext{with} H_{\mu\nu}&= \langle\phi_\mu|H[\rho_0]|\phi_\nu\rangle +
   \hal S_{\mu\nu}\sum_{C}(\gamma_{AC}+\gamma_{BC})\Delta q_C. \nonumber\\ &=
   H_{\mu\nu}^0 + H_{\mu\nu}^1; \quad \forall\, \mu\in A;\nu\in B.\label{ham}
\end{align} 
\end{subequations}
In zeroth order the Hamiltonian reduces to the first term in Eq.~(\ref{ham})
and depends solely on the reference density $\rho_0$. For systems in the
excited state or in charged or heteronuclear structures, the electronic
density differs significantly from a simple superposition of atomic ground
state densities. To a certain extent this difference is already captured at
the zeroth order level, since the coefficients which solve Eq.~(\ref{SCCb})
correspond to a time-dependent density different from $\rho_0$. This is
similar to the situation in empirical tight-binding schemes for the ground
state, where even certain ionic crystals are suffciently well described
\cite{Pola88}.  However, a consideration of the full Hamiltonian in
Eq.~(\ref{ham}) leads obviously to a more balanced treatment, because
dynamical changes in the electron density are explicitly included in a
selfconsistent fashion. In this way, one can even hope to correctly describe
the large amplitude motion induced by intense laser fields.

Solution of Eq.~(\ref{SCCb}) requires an iterative procedure with timesteps in
the attosecond regime. A symplectic algorithm is used for this task
\cite{mydiss}, which is based on the Cayley representation of the time evolution
operator and conserves the norm of the wavefunction exactly. Besides the
Hamiltonian and overlap matrices (see Ref.~\cite{Fra02} for details of the
construction) , Eq.~(\ref{SCCb}) contains also the nonadiabatic coupling
matrix $\langle\phi_\mu|\frac{\pa}{\pa \Ra}\phi_\nu\rangle$, in which all
on-site elements ($\phi_\mu, \phi_\nu$ on the same atom) are set to zero
\cite{Kue95}. This allows one to relate the remaining elements to a simple
derivative of the overlap matrix.

Variation of the Lagrangian with respect to the nuclear coordinates leads to
the following expression for the forces:
   \begin{equation}
\label{SCCf}
  \begin{split}
  M_A \ddot{{\bf R}}_A = &- \sum_i^{\text{occ}} \sum_{\mu\nu} b_{\mu i}^*
   b_{\nu i}\left( \frac{ d H_{\mu\nu}^0}{d {\bf R}_A } + \frac{d
   S_{\mu\nu}}{d {\bf R}_A } \sum_B \gamma_{AB} \Delta q_B\right)\\ &+
   \frac{1}{2} \sum_i^\text{occ} \sum_{\mu\nu\delta\gamma}\left( b_{\mu i}^*
   \frac{d S_{\mu\nu}}{d {\bf R}_A } S^{-1}_{\nu\delta} H_{\delta\gamma}
   b_{\gamma i} + \text{c.c.}  \right)\\ &-\Delta q_A \sum_{B}
   \frac{d\gamma_{AB}}{d {\bf R}_A } \Delta q_B - \frac{d E_{\text rep}}{d
   {\bf R}_A }.
\end{split}
\end{equation}
Since the Hamiltonian is time dependent due to an external field or nuclear
motion, the molecular orbital coefficients will in general represent a
coherent superposition of different eigenstates of the system. In this case,
the nuclei move in a mean potential according to Eq.~\ref{SCCf}, rather than
being restricted to a particular Born-Oppenheimer surface as in conventional
adiabatic MD approaches. In fact, due to the coupling of the EOM, energy can
be freely exchanged between the electronic and ionic subsystems as long as the
total energy of the system is conserved. Equations of motion that are
equivalent to the ones reported here, have been derived earlier by Saalmann
and Schmidt \cite{Saa96} as well as Todorov \cite{Tod01}. Including the second
order correction, they are solved here for the first time in an actual
calculation.

Todorov pointed out, that for an incomplete basis the force equation has to be
augmented by additional velocity dependent terms, which should become
important in high energy collisions. Interestingly, omission of these terms
does not violate energy but only momentum conservation. Hence, it is useful to
monitor the total momentum (electronic + ionic) of the system if
Eq.~(\ref{SCCf}) is used, as in every practical calculation the basis set is
incomplete.

Finally, in order to simulate the interaction with electromagnetic fields, the
vector potential ${\bf A}\rtb$ needs to be incorporated, which is done via
minimal coupling, ${\bf p} - \frac{e}{c}{\bf A}$. A numerically efficient
approximation was proposed by Graf and Vogl \cite{Gra95} and later by Allen
\cite{All94} and is given by:
  \begin{eqnarray}
  \label{PEIvogel}
   H_{\mu\nu}[\ro,{\bf p},{\bf A}(t)] &=& \exp{\left[ \frac{ie}{\hbar c} ({\bf
     R}_A - {\bf R}_B){\bf A}(t)\right]} \nonumber\\&& \times
     \,H_{\mu\nu}[\ro,{\bf p}];\quad \mu\in A,\nu \in B,
\end{eqnarray}
which relates the desired field dependent Hamiltonian to the already known
matrix elements of the unperturbed one. Expression (\ref{PEIvogel}) was
derived under the assumption that the radiation wavelength is much larger than
the molecular system under study, which is usually well fulfilled for
frequencies in the optical range.  It should be mentioned, that
Eq.~(\ref{PEIvogel}) in principle holds for arbitrarily strong fields in
contrast to the electric dipole approximation.

If the interest is just on calculating optical spectra, rather than the
molecular motion initiated by a laser pulse, only Eq.~(\ref{SCCb}) needs to be
solved for a fixed geometry. Following the approach of Yabana and Bertsch
\cite{Yab96,Mar03} the field in Eq.~(\ref{PEIvogel}) is turned on only at a
certain instant of time, which populates the complete manifold of excited
states. The time dependent dipole moment $d(t)$ can then be used to calculate
the dynamic polarizability:
\begin{equation}
  \label{pola}
  \alpha(\omega) = \frac{\hbar c}{e A} \int e^{i\omega t} \left( d(t) - d(0)
  \right) dt,
\end{equation}
and the dipole strength function $S(\omega)= 2\omega/\pi\, \Im \alpha(\omega)$
of the system; a quantity which can be directly compared to experimental
spectra.
\section{Applications}
\subsection{Optical spectrum of trans-butadiene}
\label{butasec}
\begin{figure} 
  \centering \rotatebox{0}{ \includegraphics[scale=1.1]{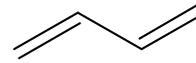}}
\caption{Schematic illustration of trans-butadiene ($C_4H_6$).\label{buta}}
\end{figure}

\begin{figure} 
  \centering \rotatebox{0}{ \includegraphics[scale=0.7]{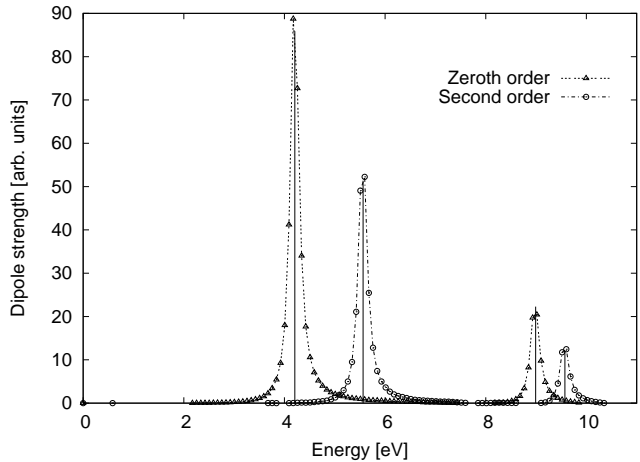}}
\caption{Dipole strength of trans-butadiene as given by the time-dependent DFTB 
  method in zeroth and second order \cite{Com2}. Shown is an average over
  different molecular orientations with respect to the polarization of the
  vector potential. Results obtained in the DFTB linear-response
  implementation of TDDFT are shown as stick spectrum. The associated
  oscillator strengths have been normalized to the maximum of the highest
  energy peak.\label{butopt}}
\end{figure}

As a first application of our method, we examine a prototypical $\pi$-system,
trans-butadiene (Fig.~\ref{buta}).
The optical spectrum of this molecule has been the subject of numerous quantum
chemical studies (see e.g.~Ref.~\onlinecite{Hsu01} and references
therein). Recently, also a detailed investigation of the molecular dynamics in
the excited state appeared \cite{Dou04}. Dou et al.~employed the DFTB method
described in this work without the second order correction. Our interest here
is to analyze the implications of including this term. To this end, we first
relaxed the molecule with the ground state DFTB method and recorded the
optical spectrum according to the prescription given in Sec.~\ref{meth}. After
applying a vector potential of A = 0.0125 gauss cm, the Kohn-Sham orbitals
were propagated for 38.7 fs with a time step of 12 as. Since the finite
sampling introduces spurious negative parts in the imaginary part of the
polarizability, the dipole moment was damped with a factor of $e^{-kt}$ (k =
0.3 eV/$\hbar$), like in Ref.~\onlinecite{Yab96}. This also simulates
dephasing or other line broadening effects which would appear more naturally
in a more complete theory. The resulting spectrum with and without second
order correction is shown in Fig.~\ref{butopt}.
As can be seen, the maximum absorption in the latter method is located at 4.21
eV. This is exactly the difference of the LUMO (lowest unoccupied molecular
orbital) and the HOMO (highest occupied molecular orbital) energies of the
ground state DFTB method. If the second order term is included in the
calculation, the absorption is strongly blue shifted to 5.56 eV, which is in
good agreement with the experimental value of 5.8 eV \cite{Doe79}. Along with
the energy shift, a reduction of absorption strength is also observed.

 To better understand the origin of these changes, we also performed
calculations with our implementation of the TDDFT linear response formalism
\cite{Nie01}. In this approach, excited state singlet energies $\omega_I$ are
given by the solution of the following eigenvalue problem:
\begin{equation}
\label{lr}
    \sum_{kl} \left[ \omega^2_{ij} \delta_{ik} \delta_{jl} + 2\sqrt{
     \omega_{ij} } K_{ij,kl} \sqrt { \omega_{kl}} \right] \; F^I_{kl} =
     \omega_I^2 \; F^I_{ij}.
\end{equation}
Here the $\omega_{ij}$ are energy differences between unoccupied orbitals $j$
and occupied orbitals $i$, while the so called coupling matrix $K$ describes
the change of the SCF potential due to the induced density. As Eq.~(\ref{lr})
shows, the effect of the coupling matrix is not only to shift the true excited
state away from simple orbital energy differences, but also to couple
different single-particle transitions. The explicit form of $K$ is given by:
 \begin{eqnarray}
   \label{coup}
    K_{ij,kl} &=& \dintp \psi_{i}\rb \psi_{j}\rb \\ &&\times\left(
    \frac{1}{|\ro-\rp|} + \frac{\delta v_{xc}[\rho]\rb}{\delta\rho\rpb}
    \right) \psi_{k}\rpb \psi_{l} \rpb\nonumber,
 \end{eqnarray}
which is nothing else than the second order term of Eq.~(\ref{LAGRAl3}), when
the induced density is expanded in particle-hole states. The results of the
linear response calculations with and without the coupling matrix contribution
are given as stick spectrum in Fig.~\ref{buta}. Obviously, there is a perfect
match beween the real time and linear response approaches to TDDFT both in the
energetical position of the states and the oscillator strength. This
equivalence had to be expected, since the linear response approach amounts to
a perturbative solution of the TDDFT equations in the small amplitude
limit. However, to our knowledge this has so far never been shown in practical
calculations.
\subsection{${\rm C}^+ - {\rm C}_{60}$ collisions}
\label{coll}
Ion-cluster collisions provide an ideal application field for approximate
TDDFT molecular dynamics simulations. This is because the number of degrees of
freedom is usually too large to be treated with first principles calculations
and also nonadiabatic effects are strong and important. Depending on the
velocity of the projectile, collisions can induce vibrational or a combination
of electronic and vibrational excitations of the cluster, where the latter
type cannot be described in conventional Born-Oppenheimer dynamics. In this
context, Kunert and Schmidt undertook a systematic investigation of
ion-fullerene collisions and provided an explanation for seemingly conflicting
experimental observations \cite{Kun01}. Their nonadiabatic quantum molecular
dynamics (NA-QMD) method is essentially equivalent to the DFTB scheme
described here in the zeroth order approximation.  Accordingly, it is
interesting to see whether the second order correction is of any benefit in
these kind of simulations.

For the special case of ${\rm C}^+ - {\rm C}_{60}$ collisions we performed
calculations for different values of the impact velocity ($v$ = 0.01 $\ldots$
0.45 a.u.) and impact parameter ($b$ =2.0,7.5 a.u.) for randomly oriented
fullerene cages. Special care had to be taken in the definition of the initial
conditions of the EOM, since the ${\rm C}^+ - {\rm C}_{60}$ configuration does
not correspond to the ground state of the system. For that reason, we
performed separate ground state calculations for the two subsystems and
combined the resulting KS orbitals to obtain the desired charge state. The
initial ion-cluster distance was chosen large enough to prevent any
interaction and the system was then left to evolve freely according to the EOM
[Eq.~(\ref{SCCb}) and (\ref{SCCf})].
\begin{figure} 
  \centering \rotatebox{0}{ \includegraphics[scale=0.7]{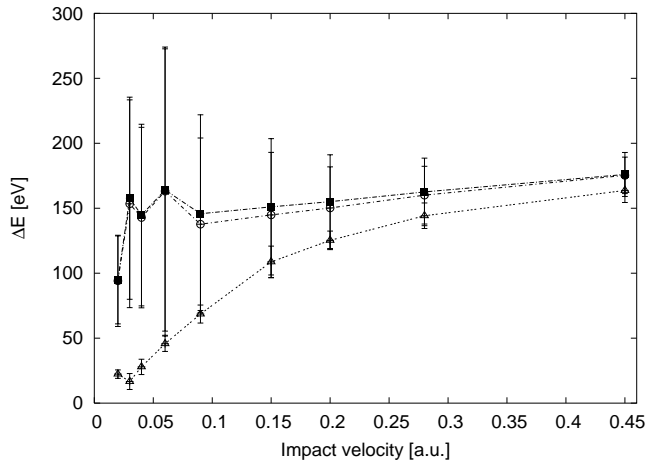}}
\caption{Total kinetic energy loss $\Delta E$ of the projectile in the center of mass
  system for an impact parameter of b = 2.0 a.u. For each velocity three
  trajectories with random cage orientation were calculated; error bars
  correspond to the standard deviation. {\em Dark squares:} second order, {\em
  open circles:} zeroth order DFTB results, {\em open triangles:} loss due to
  electronic excitation, obtained from the difference of time-dependent and
  ground state energy in the zeroth order DFTB scheme.\label{tkel}}
\end{figure}

Fig.~\ref{tkel} depicts the total kinetic energy loss $\Delta E$ in the center
of mass system that the projectile experiences due to the collision. It can be
directly compared to the results in Fig. 3 of Ref.~\onlinecite{Kun01}, which
were obtained for a single fixed collision geometry. As already shown there,
the vibrational excitation of the fullerene dominates for smaller velocities
($v < 0.1$ a.u.), while mostly electronic excitation is responsible for the
energy loss at larger impact energies. As the impact velocity increases,
$\Delta E$ first rises, peaks around 0.05 a.u and shows a weak increase beyond
0.1 a.u.~in our calculations. This is in variance with the results of Kunert
and Schmidt \cite{Kun01}, which claim velocity-independent excitation energies
in the high velocity range. Taking the strong dependence on the collision
geometry into account, an extensive phase space sampling would be necessary to
resolve this issue, which is outside the scope of this work.
\begin{figure} 
  \centering \rotatebox{0}{ \includegraphics[scale=0.7]{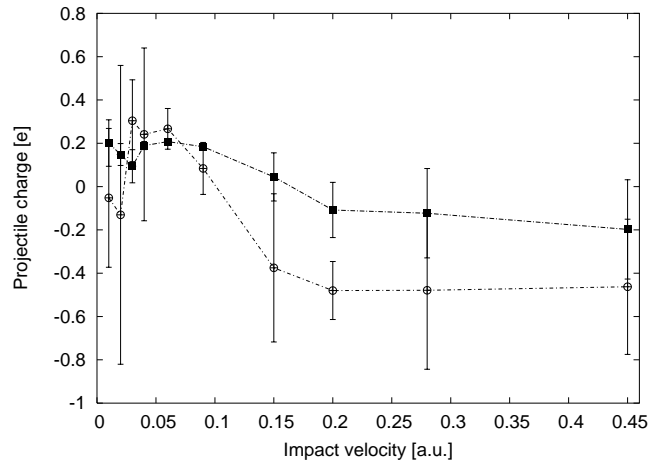}}
\caption{Charge of the carbon atom after the collision with ${\rm C}_{60}$ for
  different values of the impact velocity and an impact parameter of b = 7.5
  a.u.~. For each velocity three trajectories with random cage orientation
  were calculated; error bars correspond to the standard deviation. {\em Dark
  squares:} second order, {\em open circles:} zeroth order DFTB
  results.\label{charge}}
\end{figure}

Turning now to a comparison of the predictions of DFTB with and without second
order correction, we find only marginal differences in the results of both
methods. Such a difference could have been expected for high impact
velocities, but with such a large amount of energy deposited in the cluster,
finer details of the electronic structure seem to have negligible influence.

The second order correction has however implications for other
observables. Fig.~\ref{charge} shows the charge of the projectile after the
collision. Here fractional charges need to be understood in the probabilistic
interpretation of quantum mechanics, since in the simulations the system
remains in a superposition of eigenstates with integer charges also
asymptotically. For higher velocities, which correspond to higher electronic
excitation as mentioned above, the charge given by the zeroth order DFTB
method is significantly more negative than the second order one. This can be
explained by a larger contribution of the asymptotic ${\rm C}^- - {\rm
C}_{60}^{2+}$ state, which in the zeroth order approximation is located only
slightly higher in energy as the initial ${\rm C}^+ - {\rm C}_{60}^{0}$
state. As the results in Tab.~\ref{asym} show, this energy ordering is in
striking contrast with the one given by the second order DFTB method and
experiment. Although calculations of charge transfer cross sections have been
performed in a zeroth order scheme \cite{Kno99,Kno00}, the results of this
section suggest, that in general a more advanced treatment is absolutely
necessary.
\begin{table}
\caption{Energy ordering of different asymptotic states of the
  singly positively charged ${\rm C}-{\rm C}_{60}$ system in eV. The DFTB
  energies with and without second order correction were obtained by separate
  calculations of the two subsystems and addition of the results.  For ${\rm
  C}_{60}$, all calculations were performed at the DFTB optimized geometry of
  the neutral species. The experimental results were obtained from measured
  ionization potentials and electron affinities \cite{Nist,Cox91,Aji90}. The
  most stable state of each method was set to zero energy. \label{asym}}
\begin{ruledtabular}
\begin{tabular}{ccccc}
{\rm C} & ${\rm C}_{60}$ & zeroth order & second order & exp.\\\colrule + & 0
& 0.0 & 2.68 & 3.66\\ 0 & + & 0.58 & 0.0 & 0.0\\ -- & 2+ & 1.16 & 10.80 &
7.14$\ldots$9.14
\end{tabular}
\end{ruledtabular}
\end{table}
\subsection{Protonated Schiff base photodynamics}
\label{reti}
\begin{figure} 
  \centering \rotatebox{0}{ \includegraphics[scale=0.4]{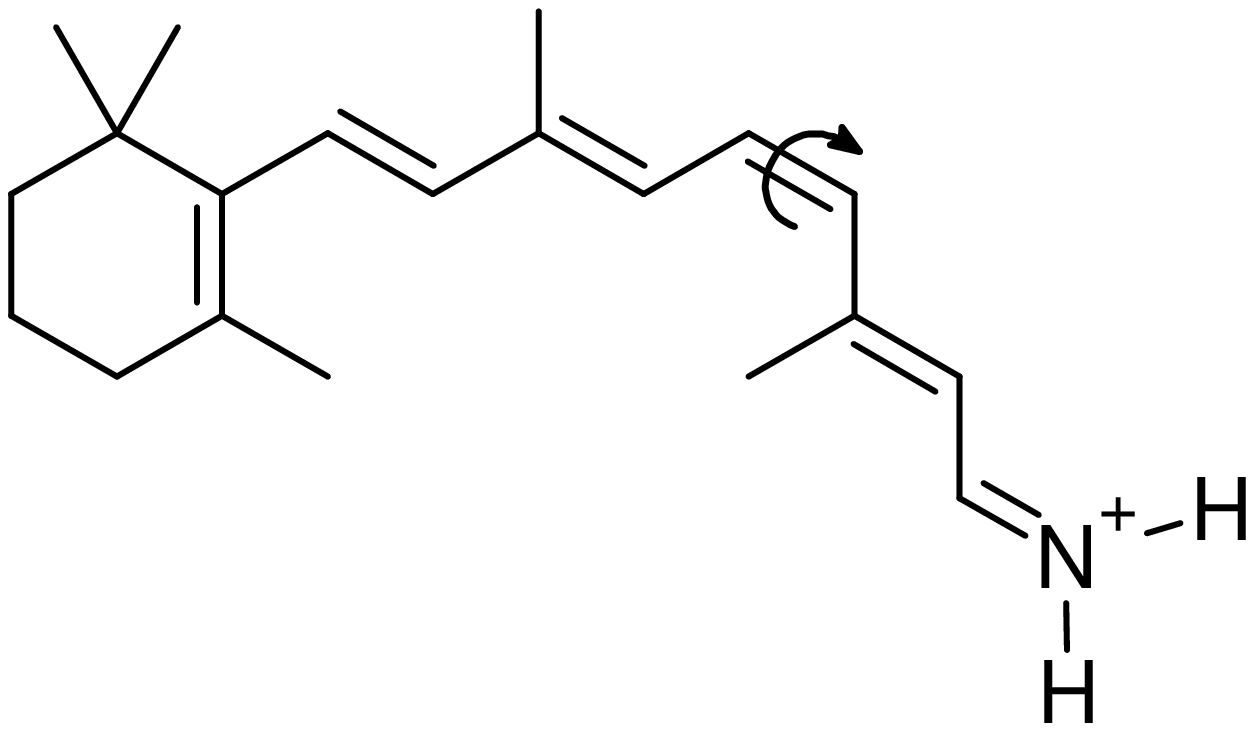}}\\
\rotatebox{0}{ \includegraphics[scale=0.4]{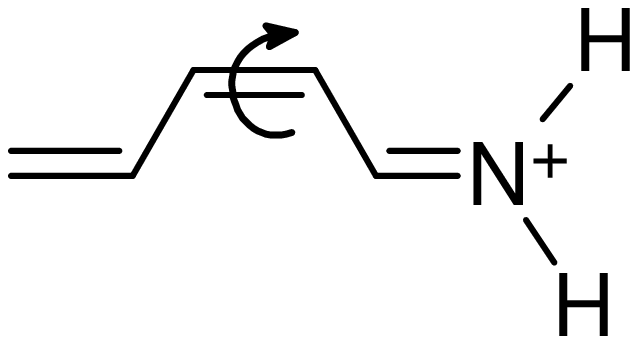}}
\caption{Top: Structure of 11-cis retinal, which is found in dark adapted
  rhodopsin and transforms to the all-trans form upon absorption of
  light. Bottom: Protonated Schiff base model used in this study.
  \label{psbstruc}} 
\end{figure}

As a last application we study the photodynamics of the retinal molecule,
which is of special importance in the field of biology. This chromophore is
found in a variety of proteins, where it initiates quite different reactions
in the cell. In bacteriorhodopsin and halorhodopsin for example, light
absorption of retinal triggers the membrane transport of protons and chloride
ions, respectively. In contrast, it starts a cascade of reactions that
initiate the vision process in rhodopsin.  In all these systems, the retinal
is known to isomerize around a specific double bond. The quantum yield of the
photoreaction is particularly high ($\Phi\approx 0.7$) and the deexcitation to
the ground state occurs in no more than 200-500 fs \cite{Aha01,Kan01}. Because
of these unusual features, this system provides an interesting subject for a
theoretical investigation. For a complete understanding of the retinal
photodynamics it would certainly be necessary to include the full protein
environment in such an investigation. However, important information can
already be drawn from the examination of small retinal analogues like the
protonated Schiff bases (Fig.~\ref{psbstruc}).
These models share a polyene chain of alternating single and double bonds with
retinal, as well as the positively charged ${\rm NH}_2^+$ Schiff base group,
which is crucial for the function of the chromophore in the protein. High
level quantum chemical CASPT2 calculations on these analogues revealed, that
after absorption the system moves out of the Franck-Condon region along the
C=C stretch normal coordinate (see Ref.~\onlinecite{Gar98} and references
therein). After inversion of single and double bonds, torsion around the
central C=C bond sets in. A barrierless path then leads to a conical
intersection of ground and excited state, where efficient deactivation to the
photoproduct occurs. In line with Stark spectroscopy \cite{Mat76,Pon83},
CASPT2 theory predicts a large charge transfer from the Schiff base group to
the other terminus upon excitation, that increases along the excited state
pathway.

Recently, we investigated the excited state potential energy surface (PES)
obtained from static DFTB calculations in the linear response formalism and
found severe deviations from the CASPT2 results \cite{Wan04}. In fact, the
only barrierless paths found to conical intersections with the ground state
involved single rather than double bond isomerization (in accordance with ab
initio TDDFT calculations). An intrinsic reaction coordinate connecting
Franck-Condon point and correct conical intersection (as described in the
CASSCF model) includes a significant barrier. Thus an efficient and ultrafast
reaction seems to be unlikely at the DFT level of theory. However, one should
keep in mind, that at finite temperature molecules posses a significant amount
of kinetic energy already at the Franck-Condon point, which allows the system
to sample a large fraction of the excited state PES. Hence, the minimum energy
path might not necessarily provide a representative description of the
photodynamical pathway. Moreover, nonadiabatic transitions are not restricted
to conical intersections. They can also occur in regions where there is a
finite gap between the ground and excited state surfaces, especially when the
atomic velocity is high. Our interest is therefore to perform nonadiabatic
molecular dynamics simulations of the PSB model system to see whether the
discrepancies between DFT predictions and experiment remain at a full
dynamical level.

A complete description of the photochemical process would in principle require
a full phase space sampling prior to excitation. Since the maximum absorption
is strongly geometry dependent, we nevertheless take only the relaxed geometry
of the ground state minimum into account. At the Franck-Condon point random
velocities corresponding to a temperature of 300 K are assigned to the atoms
and the system is left to evolve freely without further constraints. The
excitation itself is induced by a Gaussian shaped laser pulse with a central
frequency of 3.45 eV, that is slightly detuned with respect to the maximum
absorption to avoid population of higher excited states close in energy. The
maximum absorption itself is located at 3.88 eV in the second order DFTB
method and agrees well with first principles TDDFT calculations \cite{Wan04}
(4.03 eV) as well as CASPT2 \cite{Gar02} (4.02 eV) and experiment (3.85 - 4.25
eV \cite{Com3}). Similar to the case of trans-butadiene, the excitation energy
is strongly underestimated at 2.66 eV, if the second order correction is not
taken into account. Since the nonadiabatic excitation process depends strongly
on the gap between ground and excited state PES, the DFTB method without
electronic selfconsistency will therefore
provide an unrealistic description of the photodynamics and will not be used
here. The duration and fluence of the laser pulse were chosen to be 5.9 fs and
5.4 mJ/cm$^2$ (A = 0.7 gauss cm). For these parameters, the total energy of
the system rises by 3.88 eV, i.e.~a one-photon transition to the first excited
state is simulated. For a total simulation time of 1.1 ps, 100 trajectories
were propagated with a timestep of 12.1 as. This guaranteed an energy
conservation of $\Delta E/E \approx 10^{-8}$. With these parameters, one
trajectory took about 22 minutes CPU time on an Intel Xeon 3.06 GHz processor.

\begin{figure} 
  \centering \rotatebox{0}{ \includegraphics[scale=0.6]{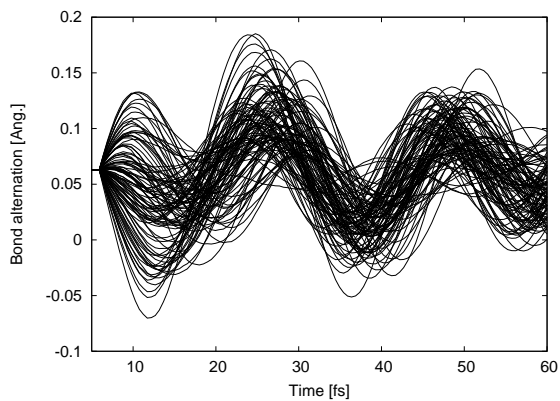}}
\caption{Bond alternation in the PSB model system for all calculated
  trajectories, estimated as the mean bond length difference between
  neighboring C-C single and double bonds.\label{bondal}}
\end{figure}

\begin{figure} 
  \centering \rotatebox{0}{ \includegraphics[scale=0.6]{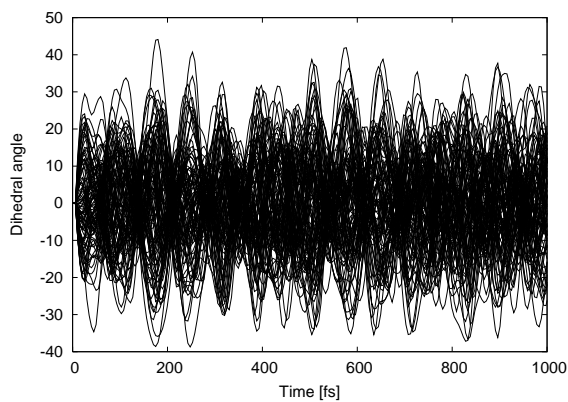}}
\caption{Torsion angle around the central double bond of the PSB model for all
  calculated trajectories.\label{dihed}}
\end{figure}

The results of the simulations show that the initial dynamics on the excited
PES are dominated by C-C stretch motions with large amplitudes up to 0.15
\AA. This is in agreement with resonant Raman studies on the PSB in solution
and in the rhodopsin protein \cite{Mat77,Lin98,Mat00}. In contrast to the
mentioned CASPT2 calculations however, the bond alternation is only inverted
for roughly half of the trajectories as shown in Fig.~\ref{bondal}.
Moreover, the time-averaged bond alternation of 0.060 \AA is only slightly
reduced with respect to the ground state minimum (0.063 \AA). Considering now
the dihedral angle which represents the torsion around the central double
bond, Fig.~\ref{dihed} shows that none of the 100 trajectories resulted in a
successful isomerization within 1 ps.
We find much larger amplitudes for the rotation around single bonds, with
torsion angles up to 80$^\circ$ for certain trajectories, although also here
no isomerization is completed in the simulation time. This preference of
single over double bond isomerization in DFT based methods was already found
in the static investigation of Ref.~\onlinecite{Wan04}. Another prediction of
CASPT2 theory, which is in nice agreement with Stark spectroscopy has already
been mentioned. The $S_0-S_1$ excitation and the subsequent motion on the
$S_1$ PES is known to involve a large charge transfer away from the Schiff
base group. We however, do not observe any significant charge transfer in our
simulations.

\begin{figure} 
  \centering \rotatebox{0}{ \includegraphics[scale=0.6]{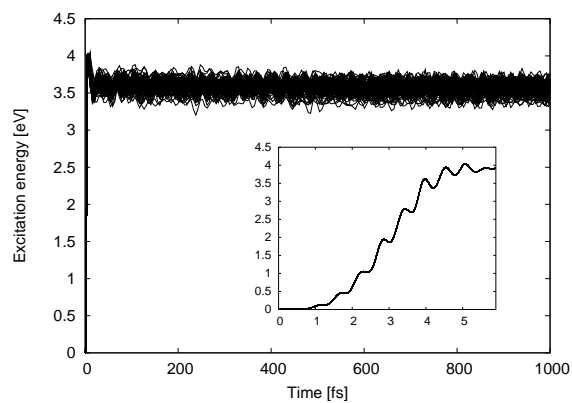}}
\caption{ Excitation energy of the PSB model versus time in eV. The inset
  shows the transition to the excited state due to the applied laser
  pulse.\label{exc}}
\end{figure}

Finally, it is interesting to analyze whether deexcitation to the ground state
occurred. In Fig.~\ref{exc} the excitation energy of the system is shown,
which is given by the difference of the time-dependent and ground state energy
at the same geometry. Directly after the end of the laser pulse, a small
reduction of the excitation energy is observed ($\approx$ 0.3 eV), which is
related to an elongation of all bonds in the PSB model. There is little change
from this point on and nonadiabatic transitions, which would manifest in a
sharp drop of the excitation energy, do not occur. In our DFT based treatment,
deactivation is therefore predicted to happen on longer timescales, presumably
involving the slower processes of internal conversion and fluorescence. At any
rate, ultrafast isomerization with a concomitant intersection of ground and
excited state is not found to be the dominant process. This is in stark
contrast to the results of Vreven et al.~\cite{Vre97}, who showed that double
bond isomerization can occur in less than 100 fs for the model at hand.

The simulations of this section could be extended by computing a larger number
of trajectories or a longer propagation time. Considering the narrow
distribution of the results presented in Fig.~\ref{bondal} to \ref{exc}, it is
not very likely that an improved sampling would reveal new
information. Extension of the simulation time might seem advantageous in light
of the experiments by Logunov et al.~\cite{Log96}, who measured an excited
state life time of 2-3 ps for the full retinal chromophore
[Fig.~\ref{psbstruc} (top)] in solution. However, for the shorter PSB model
used in this study, the excited state PES has been found to be much steeper
\cite{Gar98} in line with the mentioned study of Vreven et al.~\cite{Vre97}.
Hence, the photochemical process should be completed in the chosen simulation
time of 1 ps.

To summarize, we find that our simulations disagree in most aspects with
CASPT2 results and experiment. At the same time we confirmed the static
investigations of the DFT based potential energy surfaces from
Ref.~\onlinecite{Wan04} by dynamical calculations. Given that, (i)
time-dependent DFTB and first principles TDDFT, as well as, (ii) TDDFT with
different exchange-correlation functionals (local, gradient corrected or
hybrid) yield qualitatively the same picture \cite{Wan04}, a correct DFT based
description of the retinal photodynamics is highly unlikely. Only very
recently, failures of TDDFT in the description of inter-molecular charge
transfer states were reported \cite{Dre03}. Exchange-correlation functionals
that address this shortcoming have also been recently proposed
\cite{Taw04,Gri04,Yan04}. It will be interesting to see whether these
developments can also remedy the problems with intra-molecular charge transfer
found here.
\section{Summary} 
In this work, we presented a mixed quantum-classical approach to simulate the
coupled dynamics of electrons and nuclei. The method is based on a second
order expansion of the TDDFT Lagrangian around a suitable reference
density. We showed that the inclusion of the second order term improves both
qualitatively and quantitatively the optical spectrum of molecules. For
trans-butadiene a strong blueshift of the absorption was observed together
with a significant reduction of oscillator strength. In this context, the
analogy with the linear response approach to TDDFT revealed that the zeroth
order treatment of the Lagrangian corresponds to an uncoupled response that
neglects collective effects. Moreover, experience with the linear response
implementation suggests that this absorption shift is quite general and
especially large for $\pi-\pi^*$ transitions. For $n-\pi^*$ excitations
however, the coupling is usually weak and realistic results might already be
achieved in the zeroth order approximation. In the simulations of high energy
collision of Sec.~\ref{coll}, there is little difference in the predictions of
both schemes when the interest is in energy transfer only. This is because the
process is dominated by vibrational rather than electronic excitation in the
regime of low energy transfer, where the different level structure could be
resolved. Considering now the charge transfer, we found important differences
for the ${\rm C}^+-{\rm C}_{60}$ system, which were attributed to the
problematic description of the asymptotic states in the zeroth order
scheme. Charge transfer played also an important role in the nonadiabatic
molecular dynamics simulations of the protonated Schiff base. In contrast to
experiment we found no isomerization. This result is a negative, but we think
important one. It should be stressed, that this failure is not introduced by
our approximations but already inherent in TDDFT itself.

\section*{Acknowledgements}
The authors thank G. Seifert and M. Wanko for fruitful discussion and careful
reading of the manuscript. This work is supported by the DFG within the
Interdiscipliniary Research Group {\em Molecular Mechanisms of Retinal Protein
Action} and by the Research Training Group GK-693 of the Paderborn Institute
for Scientific Computation (PaSCo). The work of B. Torralva was performed
under the auspices of the U. S. Department of Energy by the University of
California, Lawrence Livermore National Laboratory under Contract
No. W-7405-Eng-48.


\begin{thebibliography}{99}
\bibitem{Com1} Here it is assumed, that spin-orbit coupling is neglected.
\bibitem{Com2} Please note that ionization, which sets in beyond 9.07 eV
  experimentally \cite{Nist}, can not be correctly described in a localized
  basis. In this context, a grid representation of the wavefunction together
  with an implementation of absorbing boundary conditions like in
  Ref. \onlinecite{Yab96} is more appropriate. We would also like to mention,
  that the energetical position of the second state around 9 eV is likely to
  be in error, since this excitation involves a transition to an unbound
  orbital at the minimal basis level. In spite of these problems, the full
  spectrum is shown in order to demonstrate the general principle.
\bibitem{Com3} The experimental values from Ref. \onlinecite{Arn79} correspond
  to similar but not completely equivalent structures. See also the discussion
  of this point in Ref.~\onlinecite{Wan04}.
\end{thebibliography}
\end{document}